\documentstyle[multicol,aps,epsfig]{revtex}

\makeatletter

\providecommand{\LyX}{L\kern-.1667em\lower.25em\hbox{Y}\kern-.125emX\@}

\makeatother

\begin{document}

\title{Superconductivity in \protect\( MgB_{2}\protect \) and \protect\( TaB_{2}\protect \)
: A Full-Potential Electronic Structure Comparison}

\author{Prabhakar P. Singh}

\address{Department of Physics, Indian Institute of Technology, Powai, Mumbai -400076,
India}
\date{\today}
\maketitle
\begin{abstract}
We present a comparison of electronic structure of \( MgB_{2} \) and \( TaB_{2} \),
the two new superconductors, as well as \( VB_{2} \), calculated using full-potential,
density-functional-based methods in \( P6/mmm \) crystal structure. Our results,
described in terms of (i) density of states (DOS), (ii) band-structure, and
(iii) the DOS and the electronic charge density in a small energy window around
the Fermi energy \( E_{F} \). In particular, the charge density around \( E_{F} \)
in \( MgB_{2} \) and \( TaB_{2} \) show striking similarity as far as the
\( B \) plane is concerned. A comparison of their band-structures, coupled
with \( l \)-character analysis, indicates that \( TaB_{2} \) has substantially
more \( p \)-character than \( VB_{2} \) along \( A-L \) and \( H-A \) directions
near \( E_{F} \). 
\end{abstract}
\begin{multicols} {2}
With the discovery of superconductivity in \( MgB_{2} \)\cite{akimitsu}, many
Boron compounds such as \( BeB_{2}, \) \( LiB_{2}, \) \( AlB_{2} \) \cite{felner} and
transition-metal borides\cite{layrovska,kaczorowski,gasparov} with \( AlB_{2} \)-structure
are being reexamined for possible superconductivity. The search for new \( MgB_{2} \)-like
superconductors is motivated not only by the desire to discover new superconductors
but also with the hope that it will be helpful in unravelling the nature of
interaction responsible for superconductivity in these materials. In this context,
the recently reported superconductivity in \( TaB_{2} \) \cite{kaczorowski}
with a superconducting transition temperature \( T_{c}\approx 9.5 \)\( K \),
if confirmed\cite{kaczorowski,gasparov}, may lead to a better understanding
of superconductivity in \( MgB_{2} \) and \( MgB_{2} \)-like superconductors
such as \( TaB_{2}. \) 

The continuing experimental \cite{budko,gray,tsuda} and theoretical \cite{hirsch,kortus,satta,liu,singh1} efforts
at characterizing and understanding the nature of superconductivity in \( MgB_{2} \)
have made substantial progress. However, a complete understanding of superconductivity
in \( MgB_{2} \), and for that matter in \( TaB_{2} \), is still lacking.
It is interesting to note that the superconducting transition temperatures in
\( MgB_{2} \) and \( TaB_{2} \) are \( 39 \)\( K \) and \( 9.5K \), respectively,
while \( VB_{2} \) has been found not to superconduct down to \( 0.42K \)
\cite{layrovska,kaczorowski}. Thus a theoretical study of those aspects of
electronic structure of \( MgB_{2}, \) \( TaB_{2} \) and \( VB_{2} \), which
could be responsible for superconductivity or lack thereof in these materials,
may provide some clues as to why \( TaB_{2} \) is superconducting but \( VB_{2} \)
is not. This, in turn, may improve our understanding of superconductivity in
\( MgB_{2} \) itself. 

In this paper we present a full-potential electronic structure study of \( TaB_{2} \)
and \( VB_{2} \) in \( P6/mmm \) crystal structure using density-functional-based
methods. For comparison with \( MgB_{2} \), we use our fully-relaxed, full-potential
electronic structure results for \( MgB_{2} \) in \( P6/mmm \) crystal structure\cite{singh1}.
We analyze our results in terms of (i) density of states (DOS), (ii) band-structure
along symmetry directions, and (iii) the DOS and the electronic charge density
in a small energy window around the Fermi energy, \( E_{F} \), in these systems. 

We have used our own full-potential program as well as LMTART package \cite{lmtart}
to calculate, self-consistently, the electronic structure of \( TaB_{2} \)
and \( VB_{2} \) using the experimental lattice constants. These calculations
have been used to study the band-structure along symmetry directions and site-
and symmetry-decomposed densities of states. For studying the charge density
in a small energy window around \( E_{F} \) we have used the Stuttgart TB LMTO
package \cite{tblmto}. The fully-relaxed, full-potential electronic structure
of \( MgB_{2} \) was calculated as described in Ref.\cite{singh1}

Based on our calculations, described below, we find that in \( TaB_{2} \),
within a small energy window around \( E_{F} \), the charge density in the
\( B \) plane is still two-dimensional and it is associated with the \( p_{x,y} \)
orbitals as is the case for \( MgB_{2} \) \cite{kortus,satta,singh1}. The
corresponding charge density in \( VB_{2} \) does not have this character.
A comparison of band-structures of \( MgB_{2} \), \( TaB_{2} \) and \( VB_{2} \)
along symmetry directions reveals that the \( \Gamma _{5} \) point is well
below \( E_{F} \) in both \( TaB_{2} \) and \( VB_{2} \). However, based
on our understanding 
of superconductivity
in \( MgB_{2} \) 
\cite{hirsch,kortus,satta,liu,singh1}, 
we find that the \( 7 \)-th band, specially around \( K \)
and \( A \) points with substantially more \( p \)-character in \( TaB_{2} \)
than in \( VB_{2} \) , may play an important role in making \( TaB_{2} \)
a superconductor.

Before describing our results in detail, we provide some of the computational
details of our calculations. The charge self-consistent full-potential LMTO
calculations were carried out with the generalized gradient approximation for
exchange-correlation of Perdew \emph{et al.} \cite{perdew} and 484 \( k \)-points
in the irreducible wedge of the Brillouin zone. The basis set used consisted
of \( s, \) \( p, \) \( d \) and \( f \) orbitals at the \( Ta \) (\( V \))
site and \( s, \) \( p \) and \( d \) orbitals at the \( B \) site. The
potential and the wavefunctions were expanded up to \( l_{max}=6 \). The input
to the tight-binding LMTO calculations, carried to charge self-consistency,
were similar to that of the full-potential calculations except for using spherically
symmetric potential and the space-filling atomic spheres. 

In Table I we give the lattice constants used for calculating the electronic
structure of \( MgB_{2} \), \( TaB_{2} \) and \( VB_{2} \). Note that the
lattice constants for \( TaB_{2} \) and \( VB_{2} \) are taken from experiments
\cite{kaczorowski}, while that for \( MgB_{2} \) are obtained theoretically
as described in Ref. \cite{singh1}. In Table I we also show the calculated plasma
frequencies and the total density of states per spin, \( n(E_{F}) \), at \( E_{F} \)
for these compounds. 

\vspace{0.3cm}
{\centering \begin{tabular}{|c|c|c|c|c|c|}
\hline 
&
\( a \) (a.u.)&
\( c \) (a.u.)&
\( \omega _{p}^{x} \) (\( eV \))&
\( \omega _{p}^{z} \) (\( eV \))&
\( n(E_{F}) \)\\
\hline 
\hline 
\( MgB_{2} \)&
5.76 (5.834)&
6.59 (6.657)&
7.04&
6.77&
4.70\\
\hline 
\( TaB_{2} \)&
5.826&
1.0522&
9.01&
9.71&
6.57\\
\hline 
\( VB_{2} \)&
5.665&
1.0196&
5.62&
6.60&
9.24\\
\hline 
\end{tabular}\par}
\vspace{0.3cm}

We show in Fig. 1 the total density of states for \( MgB_{2} \), \( TaB_{2} \)
and \( VB_{2} \) calculated using the full-potential LMTO method at the lattice
constants given in Table I. The gross features of the DOS of \( TaB_{2} \)
and \( VB_{2} \) are similar if one takes into account the differences arising
out of the change in the \( d \)-bandwidth as one goes from \( V \) to \( Ta \).
The comparison with \( MgB_{2} \) must be done keeping in mind the differences
in the atomic electronic configurations between \( Mg \) (\( 3s^{2} \)) and
\( Ta \) (\( 5d^{3}6s^{2} \)) and \( V \) (\( 3d^{3}4s^{2} \)). As a result
the \( d \)-DOS at \( E_{F} \) changes from \( 0.9 \) \( st/Ry \) at the
\( Mg \) site to \( 8.8 \) \( st/Ry \) at the \( Ta \) site, and finally
to \( 16.8 \) \( st/Ry \) at the \( V \) site in the respective diborides.
More importantly, we find that the \( p \)-DOS at the \( B \) site in \( MgB_{2} \),
\( TaB_{2} \) and \( VB_{2} \) decreases steadily from \( 3.36 \) \( st/Ry \)
in \( MgB_{2} \) to \( 1.55 \) \( st/Ry \) in \( TaB_{2} \) and to \( 0.58 \)
\( st/Ry \) in \( VB_{2} \). The decrease in the availability of \( p \)-electrons
around \( E_{F} \) may result in the decrease of \( T_{c} \) in \( TaB_{2} \)
and no superconductivity in \( VB_{2} \). The DOS of \( TaB_{2} \) compares
reasonably well with the cluster calculations of \cite{kawanowa}, however,
the location of \( E_{F} \) is clearly different.

In the conventional, BCS-type superconductors the DOS within an interval of
\( \pm  \)\( hw_{D} \) (\( w_{D} \)\( \equiv  \)Debye frequency) at \( E_{F} \)
is crucial for superconductivity. Since the \( B \) partial DOS is more relevant for
superconductivity in the diborides, we have plotted in Fig. 2 the symmetry-decomposed
DOS at the \( B \) site within a small energy interval around \( E_{F} \)
in \( MgB_{2} \), \( TaB_{2} \) and \( VB_{2} \). From Fig. 2 we find that
in \( TaB_{2} \) the DOS at \( E_{F} \) with \( x \) (\( y \)) symmetry
is half of the corresponding DOS in \( MgB_{2} \), while in \( VB_{2} \) it
is more than a factor of four less than in \( MgB_{2} \). Since the electrons
in the \( B \) plane are expected to play a crucial role in determining superconducting
properties, it is not surprising to find significant differences in the DOS
with \( x \) (\( y \)) symmetry in \( MgB_{2} \), \( TaB_{2} \) and \( VB_{2} \).

To study the differences in the band-structures of \( MgB_{2} \) and the transition
metal diborides as well as between \( TaB_{2} \) and \( VB_{2} \), in Fig.
3 we have plotted the band-structures along the symmetry directions for \( TaB_{2} \)
and \( VB_{2} \). The presence of \( d \)-electrons in the transition metal
diborides implies more \( d \)-like bands below \( E_{F} \) in comparison
to \( MgB_{2} \). As a result, the \( \Gamma _{5} \) point is well below \( E_{F} \)
in both \( TaB_{2} \) and \( VB_{2} \). However, the \( 7 \)-th band at \( K \)
point, with significant \( p_{z} \) character, in \( TaB_{2} \) is below \( E_{F} \)
while it is above \( E_{F} \) in \( VB_{2} \). Our \( l \)-character analysis
also shows that the \( 7 \)-th band, which crosses \( E_{F} \) along \( A-L \)
and \( H-A \) directions, has substantial \( p_{x} \) and \( p_{z} \) characters
in \( TaB_{2} \) but only a very small \( p \) character in \( VB_{2} \).
Based on our understanding of superconductivity in \( MgB_{2} \) \cite{hirsch,kortus},
it seems reasonable to state that the \( 7 \)-th band, specially around \( K \)
and \( A \) points, may play an important role in making \( TaB_{2} \) a superconductor.

To further illustrate the differences between \( MgB_{2} \),
\( TaB_{2} \) and \( VB_{2} \), we have calculated their electronic charge
densities within a \( 5 \) \( mRy \) energy window around \( E_{F} \) using
the TB LMTO method with the lattice constants as given in Table I. In these
calculations the atomic sphere radii were adjusted to minimize the discontinuity
in the Hartree potential across the atomic spheres which have been found to
give reliable results \cite{singh2}. In Fig. 4, we show the charge density
calculated within a \( 5 \) \( mRy \) energy window around \( E_{F} \) for
\( TaB_{2} \) and \( VB_{2} \). By comparing the charge density of \( TaB_{2} \)
with that of \( MgB_{2} \) as given in Ref. \cite{singh1}, we find that to
some extent the two-dimensional character of the charge density due to \( p_{x,y} \)
orbitals in the \( B \) plane remains in tact in \( TaB_{2} \). In \( VB_{2} \),
however, the decrease in the lattice constants \( a \) and \( c \) has completely
destroyed the two-dimensional nature of the charge density in the \( B \) plane,
These differences may account for no superconductivity in \( VB_{2.} \) 

In conclusion, we have presented the results of our full-potential electronic
structure calculations for \( MgB_{2} \), \( TaB_{2} \) and \( VB_{2} \)
in \( P6/mmm \) crystal structure using density-functional-based methods. We
have analyzed our results in terms of the density of states, band-structure
along symmetry directions, and the DOS and the electronic charge density in
a small energy window around the Fermi energy in these systems. In \( TaB_{2} \),
we find that within a small energy window around \( E_{F} \), the charge density
in the \( B \) plane is still two-dimensional and it is associated with the
\( p_{x,y} \) orbitals as is the case for \( MgB_{2} \). The corresponding
charge density in \( VB_{2} \) does not have this character. By comparing their
band-structures we find that the \( 7 \)-th band, specially around \( K \)
and \( A \) points with substantially more \( p \)-character in \( TaB_{2} \)
than in \( VB_{2} \), may play an important role in making \( TaB_{2} \) a
superconductor.

The author would like to thank A. V. Mahajan for helpful discussions.


\begin{figure}
\centerline{
\epsfig{file=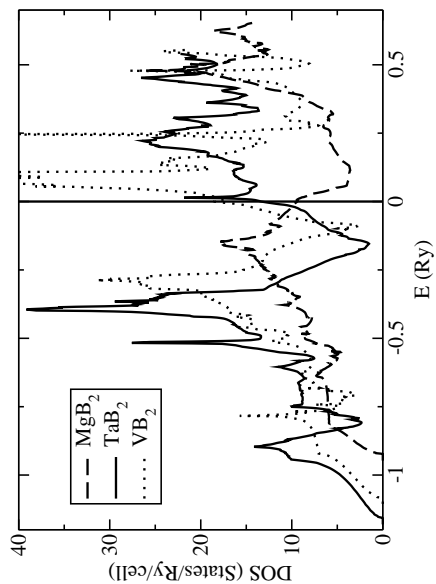,width=0.8\linewidth,clip=true}}
\caption{The total density of states calculated at the lattice
constants as given in Table I using the full-potential LMTO method. \label{fig.dos}
}
\end{figure}

\begin{figure}
\centerline{
\epsfig{file=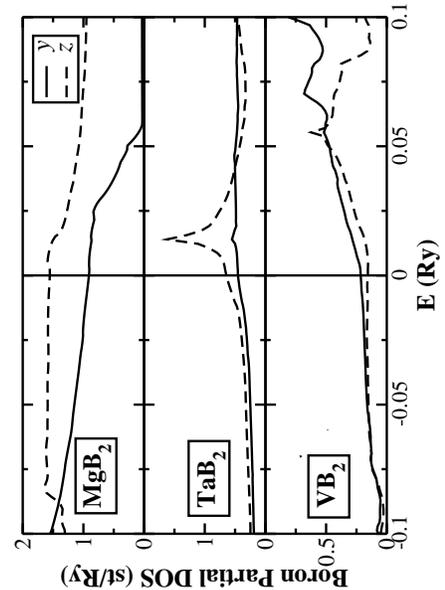,width=0.8\linewidth,clip=true}}
\caption{The partial $B$ density of states around Fermi energy
calculated at the lattice constants as given in Table I using the full-potential
LMTO method. \label{fig.dosef} } 
\end{figure}
\begin{figure}
\centerline{
\epsfig{file=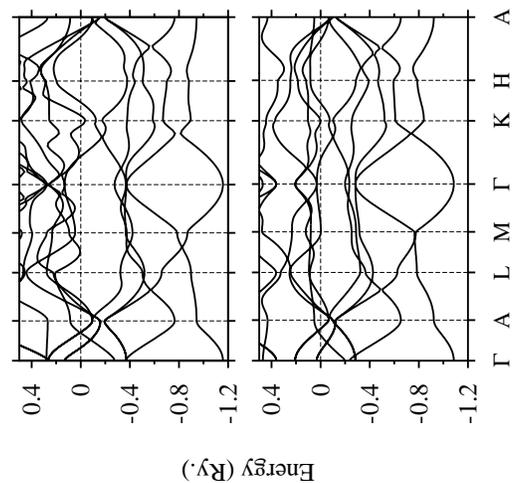,width=0.9\linewidth,clip=true}}
\caption{The
band-structure along symmetry directions calculated at the lattice constants
as given in Table I using the full-potential LMTO method. \label{bands}
} 
\end{figure}
\begin{figure}
\centerline{
\epsfig{file=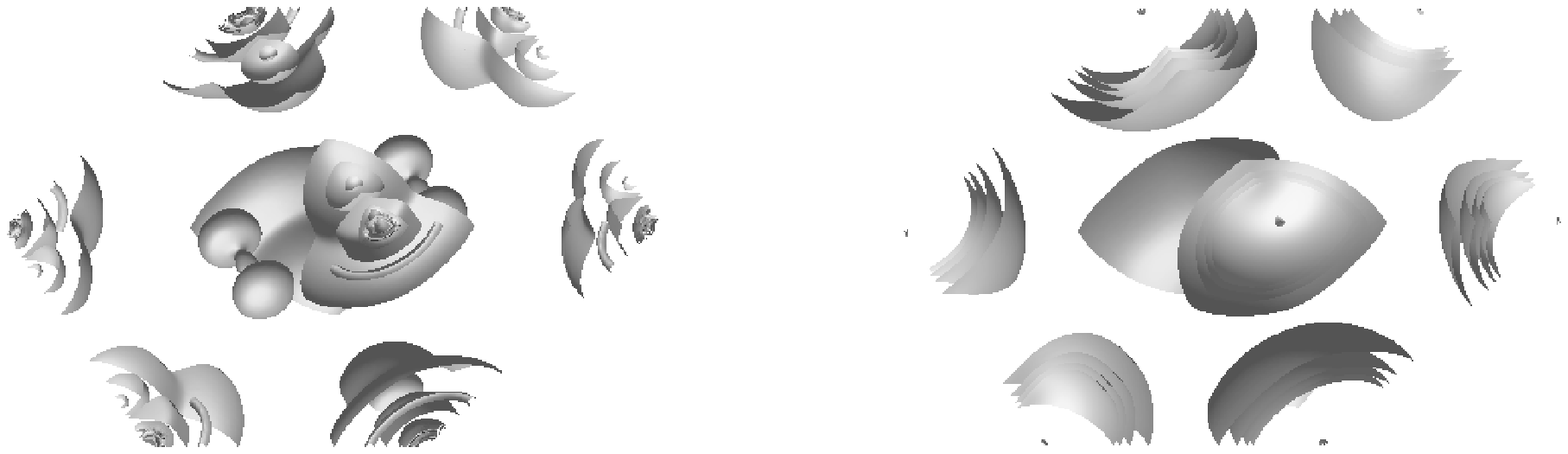,width=1.0\linewidth,clip=true}}
\caption{The isosurfaces of charge density within a $5$
$mRy$ energy window around the Fermi energy in the primitive cell of $TaB_2$
(left) and  $VB_2$ (right), calculated at the lattice constants as given in Table
I using the TB LMTO method. The values of the isosurfaces (8,6,4,2,1) decrease
as one moves away from the $B$ plane. The unit for charge density in the volume
is $10^{-4}[e/(a.u.)^3]$ and the view
is off-diagonal at an angle of $30^0$. \label{charge}
} 
\end{figure}
\end{multicols}

\begin{thebibliography}{10}
\bibitem{akimitsu}J. Akimitsu, Symp. on Transition Metal Oxides, Sendai, January 10, 2001; J.
Nagamatsu, \emph{et al}., Nature \textbf{410}, 63 (2001). 
\bibitem{felner}I. Felner, cond-mat/0102508; D. P. Young \emph{et al}., cond-mat/0104063.
\bibitem{layrovska}L. Layarovska and E. Layarovski, J. Less Common Met. \textbf{67}, 249 (1979).
\bibitem{kaczorowski}D. Kaczorowski \emph{et al}., cond-mat/0103571; D. Kaczorowski \emph{et al}.,
cond-mat/0104479.
\bibitem{gasparov}V. A. Gasparov \emph{et al.}, cond-mat/0104323.
\bibitem{budko}S. L. Bud'ko \emph{et al}., Phys. Rev. Lett. \textbf{86}, 1877 (2001). 
\bibitem{gray}H. Schmidt \emph{et al}., cond-mat/0102389; G. Karapetrov \emph{et al}., cond-mat/0102312. 
\bibitem{tsuda}Tsuda \emph{et al.}, cond-mat/0104489.
\bibitem{hirsch}J. E. Hirsch, cond-mat/0102115; J. M. An and W. E. Pickett, cond-mat/0102391.
\bibitem{kortus}J. Kortus \emph{et al}., cond-mat/0101446; K. D. Belashchenko \emph{et al}.,
cond-mat/0102290. 
\bibitem{satta}G. Satta \emph{et al}., cond-mat/0102358; Y. Kong \emph{et al}., cond-mat/0102499 
\bibitem{liu}A. Y. Liu \emph{et al.}, cond-mat/0103570.
\bibitem{singh1}Prabhakar P. Singh, Submitted to Phys. Rev. Lett.
\bibitem{lmtart}S. Y. Savrasov, Phys. Rev. B \textbf{54}, 16470 (1996). 
\bibitem{tblmto}The Stuttgart TB LMTO 4.7 package. 
\bibitem{perdew}J. P. Perdew and Y. Wang, Phys. Rev. B \textbf{45}, 13244 (1992); J. Perdew
\emph{et al}., Phys. Rev. Lett. \textbf{77}, 3865 (1996). 
\bibitem{kawanowa}H. Kawanowa \emph{et al}., Phys. Rev. B \textbf{60}, 2855 (1999).
\bibitem{singh2}Prabhakar P. Singh and A. Gonis, Phys. Rev. B \textbf{49}, 1642 (1994).
\end{thebibliography}
\end{document}